\documentstyle[preprint,aps]{revtex}
\begin{document}
\title{DIVERGENCE OF THE $~\frac{1}{N_f}~$-~SERIES EXPASION IN QED}
\author{M. Azam}
\address{Theoretical Physics Division, Central Complex,
         Bhabha Atomic Research Centre\\
         Trombay,Mumbai-400085,India}
\maketitle
\vskip .8 in
\begin{abstract}
The perturbative expansion series in coupling constant in QED
is divergent.
It is either an asymptotic series or an arrangement of a
conditionally convergent series. The sums of these types of series
depend on the way we arrange the partial sums for successive
approximations.
The $1/N_f~$ series expansion in QED, where $N_f$ is
the number of flavours, defines a
rearrangement of the perturbative series in coupling constant,
and therefore, its convergence
would serve as a proof that the perturbative series is, in fact, conditionally
convergent.Unfortunately, the $1/N_f~$ series also diverges.We proof this using
arguments similar to those of Dyson.
\par We expect that some of  the ideas and techniques discussed in
our paper will find some use in finding the true nature of the
perturbation series
in coupling constant as well as the $1/N_f~$ expansion series.
\end{abstract}
\newpage
It is more than half a century since Dyson proved that the perturbation
theory in the coupling constant in Quantum
Electrodynamics (QED) is divergent \cite{dyson}.
In this paper, we consider QED with large number of flavours
(i.e.,large number of species ) of fermions.Large flavour limit is used
in QED to argue for the existence of Landau singularity in the leading order
in $\frac{1}{N_f}$ \cite{azam,land}. Large flavour
expansion is  also used in other relativistic as well as non-relativistic
field theories.In particular, it has been very successfully used
to proof many seminal results in Landau Fermi liquid theory, again,
in the leading order in $\frac{1}{N_f}$ \cite{fel,shan,fro}.
Therefore, if the
leading order resluts are to be taken on their face values, it is imperative
to know whether the series obtained by  $\frac{1}{N_f}$~ expansion
is convergent.
\par There is another important reason why one should look for
expansion series in parameters other than the coupling constant.
We know that, even though the ordinary perturbation theory diverges,
order by order summation of the series gives excellent agreement with
experiemnts.However, we do believe that, at some order, this is going
to fail. This opens up two possibilities: either series is conditionally
convergent or it is an asymptotic power series . We explain the difference
between the two types of series.
Let us consider the series given by,
\begin{eqnarray}
~1-\frac{1}{2}+\frac{1}{3}-\frac{1}{4}+\frac{1}{5}-....~....~....
\end{eqnarray}
Term by term  summation of the series up to a
certain number of terms,
gives a good approximation to $ln2~$. However, after that,
it starts diverging.
Note that this series is an arrangement of a
conditionally convergent
series (CCS). An infinite real series is called conditionally convergent
if it converges but not absolutely.The convergence is associated with
an arrangement of the series.In other words, one can always arrange the
partial sums in such a that the series converges.Note that for a convergent
series the sum  does not dependent on the arrangement of the partial sums.
Let a CCS be given by,
$S=~\sum_{n}a_n$, then the
following property holds:(1) $\lim_{n\rightarrow\infty} a_n\rightarrow~0$
( for divergent series appearing in physical problems, it is always
hard to decide which is a generic term because terms
can always be regroupped
to define a completely different generic term ),
(2) the absolute series $~\sum_{n}|a_n|$ diverges, and (3) the negative
and positive series diverge independently. The most important property
of CCS in the present context is its behaviour under rearrangement.
It is well known that the sum of a CCS crucially depends on the way partial
sums are arranged. For example, by suitably arranging the partial sums ,
the sum of series given by Eq.(1) can be made zero.A remarkable theorem
theorem due to Riemann \cite{gelbaum,azam1} brings out this property.
\vskip .1 in
Theorem: For any given number on the real line ( including
$-\infty$ and  $+\infty$), there exists
an arrangement of a CCS such that the sum of the series converges
to it.
\vskip .1 in
There is another type of series which behaves like a convergent
series upto a certain number of terms but after that it behaves like
a divergent series. This type of series is called asymptotic series
\cite{siro} and is generally defined through a
power series represetation of a function.
Function $f(x)$ is said to have an asymptotic power series
represetation if for all $n$,
\begin{eqnarray}
\lim_{x\rightarrow 0}|\frac{f(x)-\sum_{i=0}^{n} a_i x^i}{x^n}|=0
\nonumber
\end{eqnarray}
In other words,
\begin{eqnarray}
f(x)=\sum_{i=0}^{n} a_i x^i +{\mathcal{O}}(x^n) \nonumber
\end{eqnarray}
This means that the error in estimating the function is of the same
order as the last term in the series.
To explain, let us consider the following function,
\begin{eqnarray}
F(x)=\int_{0}^{\infty}\frac{e^{-t}}{1+xt}dt
\nonumber
\end{eqnarray}
for real positive $x$ and $x\rightarrow 0$.
Since,
\begin{eqnarray}
\frac{1}{1+xt}=1-xt+x^2 t^2+...+\frac{(-xt)^k}{1+xt}
\nonumber
\end{eqnarray}
we have,
\begin{eqnarray}
F(x)=\sum_{k=0}^{N} (-1)^k x^k k! +R_{N+1}(x)~~;~~|R_N(x)|=N! x^N
\nonumber
\end{eqnarray}
The ratio of the  two successive terms is
\begin{eqnarray}
\frac{x^k k!}{x^{(k-1)} (k-1)!}=xk
\nonumber
\end{eqnarray}
This shows that the terms first decrease (since by assumption
$0<x<<1$) and then increases (when $k>\frac{1}{x})$.
From this it follows that for a given value of $x$, there exists
a best approximation. In other words, for a fixed value of
$x$, only a definite accuracy can be achieved.
These properties are found to hold for most asymptotic expansions
which appear in physical problems.
It is clear from above that under rearrangement
this type of series would be drastically altered.

\par In this paper, we consider QED with large number of flavours, $N_f$,
of fermions. Coupling constant $e$ is small,  $N_f$  is large and
$e^2 N_f$ is also small.This is what we represent by saying that
$e^2 N_f=~small~constant$ when $N_f\rightarrow\infty$.
To calculate the value any physical observable,
we can, in principle, carry out the usual perturbative analysis.
The results would clearly depend on coupling constant as well as $N_f$.
On the other hand,
we may choose to carry out expansion in $1/N_f$.This,
$1/N_f$-series expansion is based on the regroupping of the parameters
($e^2 N_f=constant$ ) and
rearrangement of the perturbative series in coupling constant.
Instead of summing the perturbative series order by order, we sum
it loop wise.First, the one loop diagrams are summed
which gives the leading term in $1/N_f$ , and then the two loop
diagrams  are summed which gives the next-to-leading term and so on.
If the original perturbative series were conditionally convergent
or an asymptotic power series,
then this rearranged series could in principle yield a different
sum of the series. We, obviously, are not in a position to calculate
the sum of the series. However, if we could argue that this series
in $1/N_f$ is convergent, then this could serve as a proof
that the perturbation series in coupling constant is conditionally
convergent. Unfortunately, it turns out that the $1/N_f$-series is
divergent. We proof this using arguments similar to those of Dyson.
\par There are claims in literature regarding the proof of the
divergence of perturbative theory based on the fact that the number
of Feynman diagrams increases factorially with the order
in the large orders of
the perturbation theory. Note that these proofs would pertain to the
absolute convergenece of the series, and the positive and negative
series independently (for a recent survey of these results see
\cite{dunne} and references there in) .  \\
Dyson's arguments for the divergence of perturbation theory in QED
is elegant in its' simplicity.
Since we will be using similar
arguments for the divergence of $~\frac{1}{N_f}~$- expansion series in QED,
we quote the following paragraphs from his paper:
"....~let
\begin{eqnarray}
F(e^2)=a_0+a_1 e^2 +a_2 e^4+...              \nonumber
\end{eqnarray}
be a physical quantity which is calculated as a formal power series in
$e^2$ by integrating the equations of motion of the theory
over a finite or infinite time.Suppose, if possible, that the series...
converges for some positive value of $e^2$; this implies that $F(e^2)$
is an analytic function of $e$ at $e=0$.Then for sufficiently small value
of $e$, $F(-e^2)$ will also be a well-behaved analytic function with
a convergent power series expansion.
\par But for $F(-e^2)$ we can also make a physical interpretation.
~...~ In the fictitious world, like charges attract each other.The potential
between static charges, in the classical limit of large distances
and large number of elementary charges, will be just the Coulomb potential
with the sign reversed.But it is clear that in the fictitious world
the vacuum state as ordinarily defined is not the state of lowest energy.
By creating a large number $N$ of electron-positron pairs, bringing the
electrons in one region of space and the positrons in
another separate region, it is easy to construct a pathological state
in which the negative potential energy of the Coulomb forces is much
greater than the total rest energy and the kinetic energy of the particles.
~......~~. Suppose that in the fictitious world the state of the
system is known at a certain time to be an ordinary physical state with
only a few particles present.There is a high  potential barrier separating
the physical state from the pathological state of equal energy; to
overcome the barrier it is necessary to supply the rest energy for
creation of many particles. Nevertheless, because of the quantum-mechanical
tunnel effect, there will always be a finite probability that in any
finite time-interval the system will find itself in a pathological state.
Thus every physical state is unstable against the spontaneous
creation of many particles.Further, a system once in a pathological state
will not remain steady; there will be rapid creation of more and more
particles, an explosive disintegration of the vacuum by spontaneous
polarization.In these circumstances it is impossible that the
integratation of the equation of motion of the theory over any finite
or infinite time interval, starting from
a given state of the fictitious world, should lead to well-defined
analytic functions.Therefore $F(-e^2)$ can not be analytic and the series
~...~ can not be convergent."
\par The central idea in Dyson's proof of the
divergence of perturbation theory in coupling constant,
as is evidient from the lenghthy quotation above,
is that the convergence of the
perturbation theory in coupling constant
would lead to the existence of pathological states
to which the normal states of QED would decay.These pathological states
correspond to states of a  quantum
field theory whose ground state or vacuum state
is unstable.In the case of~ $\frac{1}{N_f}~$ expansion series of QED,
we will proof that its' convergence also leads to the existence
of pathological states to which normal states of large flavour
QED would decay. We explicitly construct the field theory to which these
pathological states correspond. This field theory with unstable vacuum state
turns out to be different from the one constructed by Dyson.It is a quantum
field theory with commuting fermions. \\
Before we discuss the divergence of $\frac{1}{N_f}$-expansion series
in QED we would like to make some remarks. Dyson's description
of the instability of ground state in QED through spontaneous
particle production, gives the impression that his arguments
regarding divergence of perturbative series applies only to
relativistic quantum field theories. Subsequently, it has
been clarified by Arkady Veinshtein \cite{ark} and others that the
arguments  applies equally well to perturbation series in quantum
quantum mechanics such as anharmonic
oscillators with quartic interaction terms ( see \cite{dunne} for details ).
The second point concerns  the
relevance of large flavour expansion in physics. Our discussion
of large flavour expansion is centered around
QED . In this case it may look a bit
artificial because the number flavours in QED is very small.
Bellow we demonstrate that there are physical situations where large
number of flavours appear very naturally and our arguments
can be trivially carried over to these cases.
We have in mind the the Feldman Model \cite{fel} of weakly interacting
electron gas (see ref.\cite{shan,fro} for details, Sec-II of
ref. \cite{azam} for a brief description of the model
and summary of main results).
This model  describes
a condensed matter Fermi system in thermal equillibrium
at some temparature $T$ (for simplicity, assume $T=0$)
and chemical potential $\mu$.
On microscopic
scale($\approx 10^{-8}$ cm), it can be described approximately
in terms of non-relativistic electrons with
short range two body interactions.
The thermodynamic quantities such as conductivity depend only on
physical properties of the system at mesoscopic length scale
($\approx 10^{-4}$ cm),
and therefore, are determined from processes involving momenta
of the order of
$\frac{k_{F}}{\lambda}$ around the Fermi surface, where
the parameter, $\lambda >>1$,  should be thought of
as a ratio of meso-to-microscopic length scale. This is
generally refered to as
the scaling limit(large $\lambda$, low frequencies) of the system.
The most important observation of Feldman et. al.
is that in the scaling
limit, systems of non-relativistic (free)
electrons in $d$ spatial dimensions behave like a system of
multi-flavoured relativistic chiral Dirac fermions
in $1+1$ dimensions.The number of flavours
$N_f \approx ~const. ~ \lambda^{d-1}$.
\par Consider a system of non-relativistic
free electrons in $d$ spatial dimensions with the Euclidean action,

\begin{eqnarray}
S_{0}(\psi^*,\psi)=\sum_{\sigma} \int d^{d+1} x \psi_{\sigma}^*(x)
(\rm{i} \partial_0-\frac{1}{2m} \Delta -\mu)\psi_{\sigma}(x)
\end{eqnarray}
The Euclidean free fermion Green's function, $ G^{0}_{\sigma \sigma'}
(x-y)$,
where $\sigma$ and $\sigma'$  are the spin indices,
$x=(t,\vec{x})$ and $y=(s,\vec{y})$,
$t$ and $s$ are imaginary times,
$t>s$, is given by,

\begin{eqnarray}
G^{0}_{\sigma \sigma'}(x-y)=\langle \psi_{\sigma}^*(x)\psi_{\sigma}(y)
\rangle_{\mu}
=-\delta_{\sigma \sigma'} \int (dk)\frac{e^{-ik_0(t-s)+i\vec{k}
(\vec{x}- \vec{y})}}{ik_0 -(\frac{k^2}{2m}-\mu)}
\end{eqnarray}
In the scaling limit, the leading contributions to
$ G^{0}_{\sigma \sigma'}(x-y)$
come from modes whose momenta are contained in a
shell $S_{F}^{(\lambda)}$ of thickness $\frac{k_{F}}{\lambda}$
around the Fermi surface $S_{F}$.
Let us introduce the new variables
$\vec{\omega}, ~p_{\parallel}, ~p_{0}$ such that $k_{F}\vec{\omega}
\in S_F$, $p_0 =k_0$ and $\vec{k}=(k_F+p_{\parallel})\vec{\omega}$.
If $\vec{k}\in S_{F}^{(\lambda)}$, then $p_{\parallel}<<k_{F}$, and
we can approximate the integrand of Eq.(3), by dropping
$p_{\parallel}^{2}$ term in the denominator.In other words,
\begin{eqnarray}
G^{0}_{\sigma \sigma'}(x-y)= \delta_{\sigma \sigma'}\int
\frac{d {\bf \sigma} (\vec{\omega})}{(2\pi)^{d-1}} k_{F}^{d-1} e^{ik_{F}
\vec{\omega}(\vec{x}-\vec{y})} G_{c}(t-s, \vec{\omega}(\vec{x}-\vec{y}))
\end{eqnarray}
where $d {\bf \sigma}(\vec{\omega})$ is the uniform measure on unit
sphere and
\begin{eqnarray}
G_{c}(t-s, \vec{\omega}(\vec{x}-\vec{y}))=-\int \frac{dp_{0}}{2\pi}
\frac{dp_{\parallel}}{2\pi} \frac{e^{-i k_0 (t-s)+ip_{\parallel}
\vec{\omega}(\vec{x}- \vec{y})}}{ip_0 -v_{F}p_{\parallel}}
\end{eqnarray}
is the Green's function of chiral Dirac fermion in $1+1$ dimension.
$v_{F}=k_{F}/m$ is the  Fermi velocity.The $\vec{\omega}$-integration
in Eq.(4) can be further approximated by replacing it with summation
over discrete directions $\vec{\omega}_{j}$ by dividing the shell
$S_{F}^{(\lambda)}$ into $N$ small boxes $B_{\vec{\omega}_{j}}, j=1,..,N$
of roughly cubical shape.The box, $B_{\vec{\omega}_{j}}$, is centered at
$\vec{\omega}_{j} \in S_F$ and has an approximate side length
$\frac{k_F}{\lambda}$.
The number of boxes $N$ is given by,
\begin{eqnarray}
N=\frac{Volume ~of ~the ~shell ~ S_{F}^{(\lambda)}}
{Volume ~of ~the ~cubical ~boxes
~ B_{\vec{\omega}_{j}}}\nonumber
=\frac{\Omega_{d-1} k_{F}^{d-1}\times \frac{k_F}{\lambda}}
{(\frac{k_F}{\lambda})^{d}}
=\Omega_{d-1} \lambda ^{d-1}
\end{eqnarray}
where $\Omega_{d-1}$ is the surface volume of unit sphere in
$d$ spatial dimensions.
The Green's function is, now, given by
\begin{eqnarray}
G^{0}_{\sigma \sigma'}(x-y)=-\delta_{\sigma \sigma'}\sum_{\vec{\omega}_{j}}
\int \frac{dp_{0}}{2\pi} \frac{dp_{\parallel}}{2\pi} \frac{p_{\perp}}
{2\pi}
\frac{e^{-ip_0(t-s)+i\vec{p} (\vec{x}- \vec{y})}}
{ip_0 -v_{F}p_{\parallel}}
\end{eqnarray}
where $\vec{p}=p_{\parallel}\vec{\omega}+\vec{p}_{\perp}$ is a vector
in $B_{\vec{\omega}_{j}}-k_{F} \vec{\omega}_{j}$ and $p_{0}\in
\mathcal{R}$.
Thus in the scaling limit, the behaviour of a $d$-dimensional
non-relastivistic free electron gas is described by
$(N_f=)~ N=\Omega_{d-1} \lambda ^{d-1}$ flavours of free
chiral Dirac fermions in $1+1$ dimensional space-time.
The weekly interacting system electrons can be described as
interacting (1+1)-dimensional chiral fermions
with large number of flavours. In the concluding section, we
provide arguements to show that our method has some interesting
consequences for this model.

\par Now we describe the central theme of the paper.
The Langrangian of QED with number of flavours, $N_{f}$, is
given by,
\begin{equation}
{\cal L} =  \sum_{j=1}^{N_f} \bar{{\psi}}^j \Big(i\gamma^{\mu} \partial_{\mu}
+ m - e \gamma^{\mu} A_{\mu}\Big) \psi^{j} + \frac{1}{4} F_{{\mu}{\nu}}^2
\end{equation}
where $ \psi^{j}$  and $\bar{{\psi}}^j$
are the   Dirac field and its' conjugate, $j$ is the flavour index,
and $ A_{\mu}$  and
$F_{{\mu}{\nu}}$ are the electromagnetic
potential  and the field strength respectively. We will, ultimately,
be considering cases with both the positive and negative sign of $N_f$, and
therefore, we introduce the notation $|N_f|=sign(N_f)\times N_f$~ for latter
convinience.
The ~$\frac{1}{N_f}$-expansion
is introduced by assuming that, in the limit
~$|N_{f}|{\rightarrow}{\infty}$, ~$e^{2}|N_{f}|~
=constant= ~\alpha^{2} ~(say)~$ .
Alternatively, instead of the Lagrangian given by Eq.(7), we may
consider the following Lagrangian,
\begin{equation}
{\cal L} =  \sum_{j=1}^{N_f} \bar{{\psi}}^j \Big(i\gamma^{\mu} \partial_{\mu}
+ m - \frac{e}{\sqrt |N_f|}~ \gamma^{\mu} A_{\mu}\Big) \psi^{i} +
\frac{1}{4} F_{{\mu}{\nu}}^2
\end{equation}
With this form of the Langrangian, it is easy to set up Feynman diagram
technique.To each photon and fermion line corresponds their usual propagator.
Each vertex contributes a factor of $\frac{e}{\sqrt{|N_f|}}~$, each fermion
loop contributes a factor of $~(-1)~$ for anticommuting
fermions and a factor of
$~N_f~$ because of summation over fermion flavours.Using these rules, it is
easy to set up $1/N_f~$ expansion series for any physical observable.
Just as in the case of perturbation theory in the coupling constant, the
expansion  in $~\frac{1}{N_f}~$ allows us to express an observable $~F~$
in the form,
\begin{eqnarray}
F(\frac{1}{N_f})=Q_{0}+\frac{1}{N_f}~Q_{1}+\frac{1}{N_{f}^{2}}~Q_{2}+... ...
\end{eqnarray}
$Q_{0}~,~Q_{1}~,~Q_{2},~... ...~~$ are some  functions of the coupling contant.
Now suppose
that the series converges for some small value of $\frac{1}{N_f}~$ ( large
value of $N_f~$ ), then the observable function
$F(\frac{1}{N_f})$ is analytic for
$\frac{1}{N_f}=0~$ ( $N_f=\infty~$ ).Therefore, we can consider a small
negative value of $~\frac{1}{N_f}~$ ( large negative value of $N_f~$ ) for
which the function is analytic and convergent.In other words, the function
$~F(\frac{1}{N_f})~$ can be analytically continued to small negative value
of $~1/N_f~$ and the series thus obtained will be convergent.
\par What is meaning of negative $N_f~$ ? Before we look for an answer
to this question, let us calculate the effective coupling constant
for positive as well as negative $~N_f~$ using the
formal $~1/N_f~$-expansion series for the two point Green's function.
The series is assumed to be convergent, and therefore, for sufficient
large $~N_f~$, one can restrict to the leading order term.The leading
order term is given by the one-loop diagrams which can
easily be eavaluated to obtain the polarization from which one can
read off the effective coupling constant. It is given by,
\begin{equation}
e_{eff}^{2}(\Lambda^2) = {e^2  \over 1- \frac{e^2 N_f}{3\pi |N_f|}
ln \frac{\Lambda^2}{m^2}}
\end{equation}
If $N_f$ is negative,
\begin{equation}
e_{eff}^{2}(\Lambda^2) = {e^2  \over 1+ \frac{e^2}{3\pi}
ln \frac{\Lambda^2}{m^2}}
\end{equation}
From the equation above, we find that in the limit
$\Lambda\rightarrow\infty~$, $e_{eff}^{2}\rightarrow 0$, when
$N_f$ is negative. Therefore, the formal theory that we obtain from the
analytical continuation of $1/N_f~$ ( ~for large $N_f$~) to the small negative
value of $~1/N_f~$, is ( at least formally ) asymptotically free.This
seem to suggest that the physical meaning of the negative sign of
$N_f$ could possibly be traced in the free theory without the interaction
term.
\par We will argue that the  choice
of negative $N_f$ for anticommuting fermions amounts to considering
commuting fermions with positive $N_f$ .
Let us consider the Langrangian
given by Eq.(7) in four dimensional Euclidean space.
The partition function is given by the following functional
integral,
\begin{eqnarray}
{\cal Z}_{ac}=\int DA(x) D\bar{{\psi}}(x) D\psi(x) exp(-\int d^{4}x{\cal L})
\end{eqnarray}
We carry out funtional integration with respect to the anticommuting
fermion fields (grassman variables) and obtain,
\begin{equation}
{\cal Z}_{ac}=\int DA(x)~ det^{N_f}(i\gamma^{\mu} \partial_{\mu}
+ m - e \gamma^{\mu} A_{\mu})
exp(- \frac{1}{4} \int d^{4}x  F_{{\mu}{\nu}}^2)
\end{equation}
However, if one considers fermion fields to be commuting variables, then
it turns out to be functional integration over complex fields, and
we obtain,
\begin{equation}
{\cal Z}_{c}=\int DA(x)~ det^{-N_f}(i\gamma^{\mu} \partial_{\mu}
+ m - e \gamma^{\mu} A_{\mu})
exp(- \frac{1}{4} \int d^{4}x  F_{{\mu}{\nu}}^2)
\end{equation}
This expression could be obtained from the previous expression,
simply by assuming that  $N_f$ is  negative.Therefore, anticommuting
fermions  with negative $N_f$ has the same partition function as
the commuting fermions with positive $N_f$.Since, physically
interesting observables can be calculated from the partition function,
our claim is that the negative flavour anticommuting fermion is
equivalent positive flavour commuting fermions.
\par This can also be argued using formal perturbation theory.
Consider the two point Green's functions for the photons using Lagrangian
given by Eq.(8) . First, we consider just one loop diagram and show how the
contribution due to flavours appears in the calculations.
There are two vertices and a fermion loop,
each vertex contributes a factor of $~\frac{e}{\sqrt{|N_f|}}~$,
the fermion loop contributes a multiplicative factor of $~(-1)~$
because the fermions anticommute
and a multiplicative factor of  $N_f$ because of summation
over flavours of the internal fermion lines.Now, if $N_f$ happens to be
negative, the factor $~(-1)~$ and  the factor $N_f~$, combines to give the
factor $~|N_f|~$.This is also the contribution if  the fermions commute and
the flavour is positive ( the factor $~ (-1)~$ is
absent for commuting fermions).
The same procedure applies for the multiloop  diagrams.
Calculation of any observable in QED, essentially amounts
to calculating a set of Feynman diagrams.
The information regarding the anticommuting nature
of the fermions in the calculation of the Feynman diagrams
enters through the multiplicative factor of $~(-1~)$ for each fermion
loop that appear in the diagram. Each such loop, as explained above
also contributes a factor of $N_f$. Therefore, when $~N_f~$ is taken
to be negative, the over all multiplicative factor becomes
$~|N_f|~$. As explained above, we would obtain the same multiplicative
factor if we treat the fermions as commuting fields and consider
$~N_f~$ to be positive . This shows that the choice
of negative $~N_f~$ for anticommuting fermions ammounts to considering
commuting fermions with positive $~N_f~$.
We argued above, on formal
grounds, that QED with anticommuting fermions and negative value
of $~N_f~$ is asymptotically free. Therefore, formally speaking,
QED with commuting fermions and positive $~N_f~$ is asymptotically free.

\par It is well known that the free field theory of commuting
fermions does not have a stable vacuum or ground state \cite{gel,nai,wein}.
From the arguments above, it then follows that the interacting theory
also can not have a stable vacuum state.All states in this theory
are pathological. All these results follow from the single assumption
that the $1/N_f$-expansion series in QED is convergent.Therefore,
the convergence of the series in QED with anticommuting
fermions would leads to the
decay of normal states to the pathological states of QED with
commuting fermions via the process of quantum mechanical tunnelling.
Therefore, for QED to be meaningful , the  series in $\frac{1}{N_f}$ ~
expansion must diverge.
\par Our main aim in this work was to find the nature of the
divergent perturbation series.
We considered the $~1/N_f~$ expansion series
which, as explained in the text, is a rearrangement of the perturbation
series in coupling constant. Had the series converged,
we could have concluded that the perturbative series is conditionally
convergent.However, we proved that this series also diverges. We are
also unable to  say anything about the nature this divergent series
in $1/N_f$. However, our method provides some insight into
nature of  $1/N_f$ expansion series in the case of
Feldman model which
describes very well the properties of weakly interacting
system of electrons. We will describe the case when the
interaction among the electrons in the
condensed matter medium is attractive.
We have argued earlier that such a system is equivalent
to a system of $(1+1)$-dimensional Dirac fermions with
number of flavour $N_f=const.~\lambda^d$. Using
large flavour expasion technique in the renormalization group
analysis of such systems it has been proven that the
perturbative ground state of such  a system is unstable \cite{azam,fro}.
This is the BCS instability. The proof is based on the assumption
that $\frac{1}{N_f}~$ expansion series converges and one can restrict
to the leading order term. Formal analytical continuation to the
negative flavour in the renormalization group analysis, leads to
effective coupling constant,
$g_{eff}=0$. Therefore, the analytically continued theory is equivalent
to a theory of free commuting $(1+1)$-dimensional Dirac fermions.
The ground state of this theory is also unstable. Instability of
ground state of the theory at both ends of analytically
continued domains is consistent with the convergence
of $\frac{1}{N_f}$-expansion series in case of weakly interacting
electron systems with attractive interactions. In the case of
repulsive  interaction among the electrons in the condensed
matter medium, it is not hard to prove that the $1/N_f$ series
expansion is divergent. However, the arguments involved in the
proof is slightly different from the one discussed in this paper.


\end{document}